\documentclass{article}
\usepackage{spconf,amsmath,graphicx}

\usepackage{bm}
\usepackage{mathrsfs}
\usepackage{amsmath} %,cases}
\usepackage{amssymb}
\usepackage{textcomp}
\usepackage{graphicx}
\usepackage{epstopdf}
\usepackage{xcolor}

\title{Study of Robust Diffusion Recursive Least Squares Algorithms with Side Information for Networked Agents}

\name{Yi Yu $^{*\dagger}$, Haiquan Zhao$^{*}$, Rodrigo C.~de Lamare $^{\dagger\ddagger}$ and Yuriy Zakharov $^\dagger$ \thanks{This work was partially supported by National Science Foundation of P.R. China (Grant: 61271340, 61571374 and 61433011), and the work of Y. Yu was also supported by China Scholarship Council Funding. The work of Y. Zakharov was partially supported by EPSPC grant EP/P017975/1.}}

\address{* \normalsize{\emph{School of Electrical Engineering, Southwest Jiaotong University, Chengdu, China}}\\
    $\dagger$ \normalsize{\emph{Department of Electronic Engineering, University of York, York YO10 5DD, U.K.}}\\
    $\ddagger$ \normalsize{\emph{CETUC, PUC-Rio, Rio de Janeiro 22451-900, Brazil}}\\
    \normalsize{yuyi\_xyuan@163.com, hqzhao\_swjtu@126.com, rcdl500@ohm.york.ac.uk, yury.zakharov@york.ac.uk}}
\begin{document}
\ninept
\maketitle

\begin{abstract}
This work develops a robust diffusion recursive least squares algorithm to mitigate the performance degradation often experienced in networks of agents in the presence of impulsive noise. This algorithm minimizes an exponentially weighted least-squares cost function subject to a time-dependent constraint on the squared norm of the intermediate estimate update at each node. With the help of side information, the constraint is recursively updated in a diffusion strategy. Moreover, a control strategy for resetting the constraint is also proposed to retain good tracking capability when the estimated parameters suddenly change. Simulations show the superiority of the proposed algorithm over previously reported techniques in various impulsive noise scenarios.
\end{abstract}
\section{Introduction}

In the last decade, distributed adaptive algorithms for estimating
parameters of interest over wireless sensor networks with multiple
nodes (or agents) have attracted significant attention, due to their
performance advantages and robustness \cite{sayed2014adaptation}.
The core idea is that each node performs adaptive estimation, in
cooperation with its neighboring nodes. Distributed adaptive
algorithms have been applied to many problems, e.g., frequency
estimation in power grid \cite{kanna2015distributed} and spectrum
estimation \cite{miller2016distributed}. According to the
cooperation strategy of interconnected nodes, existing algorithms
can be categorized as the incremental \cite{li2010distributed},
consensus \cite{tu2012diffusion}, and diffusion
\cite{lopes2008diffusion, cattivelli2008diffusion,chen2012diffusion}
types. The diffusion type is the most popular
\cite{tu2012diffusion}, because it does not require a Hamiltonian
cycle path as does the incremental type \cite{li2010distributed}; it
is stable and has a better estimation performance than the consensus
type \cite{tu2012diffusion}. Several diffusion-based distributed
algorithms have been proposed such as the diffusion least mean
square (dLMS) algorithm \cite{lopes2008diffusion}, diffusion
conjugate gradient (dCG) \cite{dcg}, diffusion recursive least
squares (dRLS) algorithm \cite{cattivelli2008diffusion}, and their
modifications
\cite{lee2015variable,xu2015adaptive,xu2015distributed,liu2014distributed,djio}.

In practice, measurements at the network nodes can be corrupted by
impulsive noise \cite{blackard1993measurements}. An impulsive noise
process has the property that its occurence probability is small and
the magnitude is typically much larger than the nominal measurement.
It is well-known that the impulsive noise deteriorates significantly
the performance of algorithms in the single-agent case. For
distributed algorithms in the multi-agent case, impulsive noise can
propagate over the entire network due to the exchange of information
among nodes. To reduce the impulsive noise interference, many robust
distributed algorithms have been proposed
\cite{ni2016diffusion,ahn2017new,al2017robust,kumar2016diffusion,sorsvd,crutv,rdrab,locsme,okspme}.
Some algorithms, e.g., the diffusion sign error LMS (dSE-LMS)
\cite{ni2016diffusion}, are based on using the instantaneous
gradient-descent method to minimize an individual robust criterion.
In \cite{ahn2017new}, a robust variable weighting coefficients dLMS
(RVWC-dLMS) algorithm was developed, which only considers the data
and intermediate estimates from nodes not affected by impulsive
noise; this is based on a judgement whether impulsive noise samples
occur or not. However, these robust algorithms have slow
convergence, especially for colored input signals.

RLS-based algorithms have a good decorrelating property for colored signals, thereby providing fast convergence. In this paper, therefore, we present a robust dRLS (R-dRLS) algorithm for distributed estimation over networks affected by impulsive noise. The R-dRLS algorithm minimizes a local exponentially weighted least-squares (LS) cost function subject to a time-dependent constraint on the squared norm of the intermediate estimate at each node. Unlike the framework in \cite{vega2009fast}, we consider here a multi-agent scenario with the diffusion strategy. Furthermore, in order to equip the R-dRLS algorithm with the ability to withstand sudden changes in the environment, we also propose a diffusion-based distributed nonstationary control (DNC) method. This paper is organized as follows. In Section 2, the estimation problem in the network is described. In Section 3, the proposed algorithm is derived. In Section 4, results of simulation in impulsive noise scenarios are presented. Finally, conclusions are given in Section 5.
%Then, we utilize the dichotomous coordinate-descent (DCD) method to compute the Kalman gain vector. The DCD is a shift-and-add technique without any multiplication operation, so the resulting DCD-R-dRLS algorithm has a low computational complexity.
\section{Problem Formulation}
Let us consider a network that has $N$ nodes distributed over some region in space, where a link between two nodes means that they can communicate directly with each other. The neighborhood of node \emph{k} is denoted by $\mathcal{N}_k$, i.e., a set of all nodes connected to node $k$ including itself. The cardinality of $\mathcal{N}_k$ is denoted by $n_k$. At every time instant $i\geq 0$, every node \emph{k} observes a data regressor vector $\bm u_{k,i}$ of size $M\times1$ and a scalar measurement $d_k(i)$, related as:
\begin{equation}
d_k(i) = \bm u_{k,i}^T \bm w^o + v_k(i),
\label{001}
\end{equation}
where the superscript $T$ denotes the transpose, $\bm w^o$ is a parameter vector of size $M\times1$, and $v_k(i)$ is the additive noise at node \emph{k}. The regressors $\bm u_{k,i}$ and $\bm u_{l,j}$ are spatially independent for $k\neq l$ and all $i,j$. The additive noises $v_k(i)$ and $v_l(j)$ are spatially and temporally independent for $k\neq l$ and $i\neq j$. Moreover, any $\bm u_{k,i}$ is independent of any $v_l(j)$. The model (\ref{001}) is widely used in many applications~\cite{sayed2014adaptation,sayed2011adaptive}.

The task is to estimate $\bm w^o$, using the available data collected at nodes, i.e., $\{\bm u_{k,i},d_k(i)\}_{k=1}^N$. For this purpose, the global LS-based estimation problem is described as \cite{cattivelli2008diffusion}:
\begin{equation}
\label{002}
\begin{array}{rcl}
\begin{aligned}
&\bm w_i = \arg \min \limits_{\bm w} \\
&  \left\lbrace  \lambda^{i+1}\delta\lVert \bm w\rVert_2^2 + \sum\limits_{j=0}^i \lambda^{i-j} \sum\limits_{k=1}^N   \left(d_k(j)-\bm u_{k,i}^T\bm w\right)^2 \right\rbrace,
\end{aligned}
\end{array}
\end{equation}
where $\lVert \cdot \rVert_2$ denotes the $l_2$-norm of a vector, $\delta >0$ is a regularization constant, and $\lambda$ is the forgetting factor. The dRLS algorithm solves \eqref{002} in a distributed manner \cite{cattivelli2008diffusion}. In practice, $v_k(i)$ may contain impulsive noise, severely corrupting the measurement $d_k(i)$. With such noise processes, the algorithms obtained from \eqref{002}, e.g., the dRLS algorithm, would fail to work.

\section{Proposed Distributed Algorithm}

\subsection{Derivation of the R-dRLS Algorithm}

We focus here on the adapt-then-combine (ATC) implementation of the
diffusion strategy, which has been shown to outperform the
combine-then-adapt (CTA) implementation\footnote{ In fact, the CTA
version is obtained by reversing the adaptation step and combination
step in the ATC version. } \cite{tu2012diffusion}. Following the
ATC-based diffusion strategy \cite{lopes2008diffusion,
cattivelli2008diffusion}, i.e., performing first the adaptation step
and then the combination step, the R-dRLS algorithm will be derived
in the sequel. We start with the adaptation step. Every node $k$, at
time instant $i$, finds an intermediate estimate $\bm \psi_{k,i}$ of
$\bm w^o$ by minimizing the individual local cost function:
\begin{equation}
\label{003}
\begin{array}{rcl}
\begin{aligned}
J_k(\bm \psi_{k,i}) =& \lVert \bm \psi_{k,i}-\bm w_{k,i-1} \rVert_{\bm Q_{k,i}}^2 \\
&+  [d_k(i)-\bm u_{k,i}^T\bm \psi_{k,i}]^2 ,
\end{aligned}
\end{array}
\end{equation}
with $\bm Q_{k,i} =\bm R_{k,i}-\bm u_{k,i}\bm u_{k,i}^T$, where
\begin{equation}
\label{004}
\begin{array}{rcl}
\begin{aligned}
\bm R_{k,i} \triangleq &\lambda^{i+1}\delta \bm I + \sum\limits_{j=0}^i \lambda^{i-j}\bm u_{k,j}\bm u_{k,j}^T\\
=& \lambda \bm R_{k,i-1} + \bm u_{k,i}\bm u_{k,i}^T
\end{aligned}
\end{array}
\end{equation}
is the time-averaged correlation matrix for the regression vector at node $k$ and $\bm w_{k,i-1}$ is an estimate of $\bm w^o$ at node $k$ at time instant $i-1$. Notice that the form $\lVert \bm x \rVert_{\bm Q}^2 \triangleq \bm x^T\bm Q\bm x$ in \eqref{003} defines the  Riemmanian distance \cite{pelekanakis2014adaptive} between vectors $\bm \psi_{k,i}$ and $\bm w_{k,i-1}$. Setting the derivative of $J_k(\bm \psi_{k,i})$ with respect to $\bm \psi_{k,i}$ to zero, we obtain
\begin{equation}
\label{005}
\begin{array}{rcl}
\begin{aligned}
\bm \psi_{k,i} = \bm w_{k,i-1} + \bm P_{k,i} \bm u_{k,i} e_k(i),\\
\end{aligned}
\end{array}
\end{equation}
where $e_k(i) = d_k(i)-\bm u_{k,i}^T \bm w_{k,i-1}$ stands for the output error at node $k$ and $\bm P_{k,i} \triangleq \bm R_{k,i}^{-1} $. Using the matrix
inversion lemma \cite{sayed2011adaptive}, we have
\begin{equation}
\label{006}
\begin{array}{rcl}
\begin{aligned}
\bm P_{k,i} = \frac{1}{\lambda} \left( \bm P_{k,i-1} - \frac{\bm P_{k,i-1}\bm u_{k,i} \bm u_{k,i}^T\bm P_{k,i-1}}{\lambda +\bm u_{k,i}^T \bm P_{k,i-1} \bm u_{k,i}} \right),\\
\end{aligned}
\end{array}
\end{equation}
where $\bm P_{k,i}$ is initialized as $\left. \bm P_{k,0}=\delta^{-1} \bm I\right. $ and $\bm I$ is an identity matrix. Since $\bm w_{k,i-1}=\bm R_{k,i-1}^{-1} \bm z_{k,i-1}$, where $\bm z_{k,i} = \lambda \bm z_{k,i-1} + \bm u_{k,i} d_k(i)$, \eqref{005} means that every node $k$ performs an RLS update. However, with the update \eqref{005}, the adverse effect of an impulsive noise sample at time instant $i$ will propagate through nodes via $e_k(i)$. This effect can last for many iterations. To make the algorithm robust in impulsive noise scenarios, we propose to minimize \eqref{003} under the following constraint:
\begin{equation}
\label{007}
\begin{array}{rcl}
\begin{aligned}
\lVert \bm \psi_{k,i}-\bm w_{k,i-1}\rVert_2^2 \leq \xi_k(i-1),
\end{aligned}
\end{array}
\end{equation}
where $\xi_k(i)$ is a positive bound. This constraint is employed to
enforce the squared norm of the update of the intermediate estimate
not to exceed the amount  $\xi_k(i-1)$ regardless of the type of
noise (possibly, impulsive noise), thereby guaranteeing the
robustness of the algorithm. If \eqref{005} satisfies \eqref{007},
i.e.,
\begin{equation}
\label{008}
\begin{array}{rcl}
\begin{aligned}
\lVert \bm g_{k,i} \rVert_2 \lvert e_k(i)\rvert \leq \sqrt{\xi_k(i-1)},
\end{aligned}
\end{array}
\end{equation}
where $\left. \bm g_{k,i}\triangleq \bm P_{k,i} \bm u_{k,i}\right. $ represents the Kalman gain vector, then \eqref{005} is a solution of the above constrained minimization problem. On the other hand, if \eqref{008} is not satisfied (usually in the case of appearance of impulsive noise samples), i.e., $\left. \lVert \bm g_{k,i} \rVert_2 \lvert e_k(i)\rvert > \sqrt{\xi_k(i-1)}\right. $, we propose to replace the update \eqref{005} by a normalized form  to satisfy the constraint \eqref{007}, which is described by
\begin{equation}
\label{009}
\begin{array}{rcl}
\begin{aligned}
\bm \psi_{k,i} = \bm w_{k,i-1} + \sqrt{\xi_k(i-1)} \frac{\bm g_{k,i}}{\lVert\bm g_{k,i}\rVert_2}\text{sign}(e_k(i)),\\
\end{aligned}
\end{array}
\end{equation}
where $\text{sign}(\cdot)$ is the sign function. Consequently, combining \eqref{005}, \eqref{008} and \eqref{009}, we obtain the adaptation step for each node $k$ as:
\begin{equation}
\label{010}
\begin{array}{rcl}
\begin{aligned}
\bm \psi_{k,i} = \bm w_{k,i-1} + \min \left[  \frac{\sqrt{\xi_k(i-1)}}{\lVert \bm g_{k,i} \rVert_2 \lvert e_k(i)\rvert},\; 1 \right] \bm g_{k,i} e_k(i).\\
\end{aligned}
\end{array}
\end{equation}
Then, at the combination step, the intermediate estimates
$\psi_{m,i}$ from the neigborhood $m\in\mathcal{N}_k$ of node $k$
are linearly weighed, yielding a more reliable estimate $\bm
w_{k,i}$ \cite{sayed2014adaptation}:
\begin{equation}
\label{011}
\begin{array}{rcl}
\begin{aligned}
\bm w_{k,i} = \sum\limits_{m\in\mathcal{N}_k}c_{m,k} \bm \psi_{m,i},
\end{aligned}
\end{array}
\end{equation}
where the combination coefficients $\{c_{m,k}\}$ are non-negative, and satisfy:
\begin{equation}
\label{012}
\sum\limits_{m\in\mathcal{N}_k}c_{m,k} =1 \text{, and } c_{m,k}=0 \text{ if } m\notin\mathcal{N}_k.
\end{equation}
Note that, $c_{m,k}$ is a weight that node $k$ assigns to the intermediate estimate $\bm \psi_{m,i}$ received from its neighbor node $m$. In general, $\{c_{m,k}\}$ are determined by a static rule (e.g., the Metropolis rule \cite{takahashi2010diffusion} that we adopt in this paper) which keeps them constant in the estimation, or an adaptive rule \cite{takahashi2010diffusion}.
%In doing so, a diffusion-based distributed update is formulated as:
%\begin{equation}
%\label{014}
%\begin{array}{rcl}
%\begin{aligned}
%\bm \psi_{k,i} =& \bm w_{k,i-1} + \min \left[  \frac{\sqrt{\xi_k(i-1)}}{\lVert \bm g_{k,i} \rVert_2 \lvert e_k(i)\rvert},\; 1 \right]  \bm g_{k,i} e_k(i),\\
%\bm w_{k,i} =& \sum\limits_{m\in\mathcal{N}_k}c_{m,k} \bm \psi_{m,i},
%\end{aligned}
%\end{array}
%\end{equation}
%where we also reformulate $e_k(i) = d_k(i)-\bm u_{k,i}^T \bm w_{k,i-1}$.
It is evident that the bound $\xi_k(i)$ controls the robustness of
the algorithm against impulsive noise and influences its dynamic
behavior, so choosing its value properly is of fundamental
importance. To this end, motivated by the single-agent case in
\cite{vega2009fast}, $\xi_k(i)$ is adjusted recursively based on the
diffusion strategy as:
\begin{equation}
\label{013}
\begin{array}{rcl}
\begin{aligned}
\zeta_k(i) =& \beta \xi_k(i-1) + (1-\beta) \left\| \bm \psi_{k,i}-\bm w_{k,i-1}\right\|_2^2 \\
=\beta  \xi_k&(i-1) + (1-\beta) \min [\lVert \bm g_{k,i}\rVert_2^2 e_k^2(i),\xi_k(i-1) ], \\
\xi_k(i) =&\sum \limits_{m\in\mathcal{N}_k}c_{m,k} \zeta_m(i),
\end{aligned}
\end{array}
\end{equation}
where $\beta$ is a forgetting factor, $0<\beta \lesssim 1$. In \eqref{013}, at every node~$k$, $\xi_k(i)$ can be initialized as $\xi_k(0)= E_c\sigma_{d,k}^2/(M\sigma_{u,k}^2)$, where $E_c$ is a positive integer, and $\sigma_{d,k}^2$ and $\sigma_{u,k}^2$ are powers of signals $d_k(i)$ and $\bm u_{k,i}$, respectively. The proposed algorithm is shown in Table \ref{table_2}.
\begin{table}[tbp]
    \scriptsize
    \centering
    \caption{ Proposed R-$\rm d$RLS Algorithm with the DNC Method.}
    \label{table_2}
    \begin{tabular}{lc}
        \hline
        \text{Parameters:} $0<\beta \lesssim 1$, $\lambda$, $\delta$ and $E_c$ (R-dRLS); $\varrho$ and $t_\text{th}$ (DNC)\\
        \text{Initialization}: $\bm w_{k,0} = \bm 0$, $\bm P_{k,0}=\delta^{-1} \bm I$ and $\xi_k(0)= E_c \frac{\sigma_{d,k}^2}{M\sigma_{u,k}^2}$ (R-dRLS);\\ \;\;\;\;\;\;\;\;\;\;\;\;\;\;\;\;\;\;\;\;$\varTheta_{\text{old},k}=\varTheta_{\text{new},k}=0$, $V_t=\varrho M$, and $V_d = 0.75V_t$ (DNC)\\
        \hline
        \textbf{R-dRLS algorithm:} \\
        $e_k(i) = d_k(i)-\bm u_{k,i}^T \bm w_{k,i-1}$ \\
        $\bm P_{k,i} = \frac{1}{\lambda} \left( \bm P_{k,i-1} - \frac{\bm P_{k,i-1}\bm u_{k,i} \bm u_{k,i}^T\bm P_{k,i-1}}{\lambda +\bm u_{k,i}^T \bm P_{k,i-1} \bm u_{k,i}} \right) $ \\
        $\bm g_{k,i} = \bm P_{k,i}\bm u_{k,i}$\\
        $\bm \psi_{k,i} = \bm w_{k,i-1} + \min \left[  \frac{\sqrt{\xi_k(i-1)}}{\lVert \bm g_{k,i} \rVert_2 \lvert e_k(i)\rvert},\; 1 \right]  \bm g_{k,i} e_k(i)$\\
        $\bm w_{k,i} = \sum\limits_{m\in\mathcal{N}_k}c_{m,k} \bm \psi_{m,i}$\\
        \textbf{DNC method:} \\
        \text{Step 1: to compute} $\varDelta_k(i)$\\
        \text{if}\;$i=nV_t, n=0,1,2,...$\\
        \;\;\;$\bm a_{k,i}^T = \mathcal{O} \left( \left[ \frac{e_k^2(i)}{\|\bm u_{k,i}\|_2^2},\frac{e_k^2(i-1)}{\|\bm u_{k,i-1}\|_2^2},\text{...}, \frac{e_k^2(i-V_t+1)}{\|\bm u_{k,i-V_t+1}\|_2^2} \right] \right) $\\
        \;\;\;$\varTheta_{\text{new},k} = \sum\limits_{m\in\mathcal{N}_k}c_{m,k} \frac{\bm a_{m,i}^T \bm e}{V_t-V_d}$\\
        \;\;\;$\varDelta_k(i) = \frac{\varTheta_{\text{new},k}-\varTheta_{\text{old},k}}{\xi_k(i-1)}$\\
        \text{end}\\
        \hline
        \text{Step 2: to reset} $\xi_k(i)$\\
        \text{if} $\varDelta_k(i)> t_\text{th} $\\
        \;\;$\zeta_k(i) = \xi_k(0)$, $\bm P_{k,i}= \bm P_{k,0}$\\
        \text{elseif} \; $\varTheta_{\text{new},k}>\varTheta_{\text{old},k}$\\
        \;\;$\zeta_k(i) = \xi_k(i-1) + (\varTheta_{\text{new},k}-\varTheta_{\text{old},k})$\\
        \text{else}\\
        \;\;$\zeta_k(i) = \beta \xi_k(i-1) + (1-\beta) \min \left[ \lVert \bm g_{k,i}\rVert_2^2 e_k^2(i),\;\xi_k(i-1) \right]$\\
        \text{end}\\
        $\xi_k(i) =\sum\limits_{m\in\mathcal{N}_k}c_{m,k} \zeta_m(i)$\\
        $\varTheta_{\text{old},k} = \varTheta_{\text{new},k}$\\
        \hline
    \end{tabular}
\end{table}

\textbf{Remark}: As can be seen from \eqref{010}, the operation mode
of the proposed algorithm is twofold. At time instant $i$, if
$\lVert \bm g_{k,i}\rVert_2^2 e_k^2(i) \leq \xi_k(i-1)$, the RLS
update is performed; if not, the RLS update is normalized to have a
norm of value $\xi_k(i-1)$. At the early iterations, the values of
$\xi_k(i)$ can be high compared to $\lVert \bm
g_{k,i}\rVert_2^2e_k^2(i)$ so that the algorithm will behave as the
dRLS algorithm, providing a fast convergence. Whenever an impulsive
noise sample appears, due to its significant magnitude, the
algorithm will work as an dRLS update multiplied by a very small
'step size' scaling factor given by $\sqrt{\xi_k(i-1)}/(\lVert \bm
g_{k,i}\rVert_2 |e_k(i)|)$, thus suppressing the negative influence
of impulsive noise on the estimation
\cite{song2013normalized,l1stap,hur2017variable,jio,jidf,sjidf,spa,mfsic,mbdf}
and reducing the error propagation effect. The algorithm robustness
to impulsive noise is further maintained, due to decreasing
$\xi_k(i)$ over the iterations. This algorithm can be considered as
an improved dRLS algorithm with an additional 'step size' scaling
factor which is time-varying and between 1 and
$\sqrt{\xi_k(i-1)}/(\lVert \bm g_{k,i}\rVert_2 |e_k(i)|)$, as can be
observed in \eqref{010}.

\subsection{DNC Method}
Although the decreasing values of the sequence $\{\xi_k(i)\}$ with the iteration $i$ prompt the R-dRLS algorithm more robust against impulsive noises, the algorithm also loses its tracking capability for a sudden change of the unknown vector $\bm w^o$. To improve the tracking capability, referring to the single-agent scenario \cite{vega2008new}, we also develop a diffusion-based DNC method, summarized in Table~\ref{table_2}. The DNC method includes two implementation procedures.

Firstly, a variable $\varDelta_k(i)$ at node $k$ is computed once for every $V_t$ iterations, to judge whether the unknown vector has a change or not. In this step, $\bm a_{k,i}^T = \mathcal{O} \left( \left[ \frac{e_k^2(i)}{\|\bm u_{k,i}\|_2^2},\frac{e_k^2(i-1)}{\|\bm u_{k,i-1}\|_2^2},\text{...}, \frac{e_k^2(i-V_t+1)}{\|\bm u_{k,i-V_t+1}\|_2^2} \right] \right)$ with $\mathcal{O}(\cdot)$ denoting the ascending arrangement for its arguments, and $\bm e=[1,...,1,0,...,0]^T$ is a vector whose first $V_t-V_d$ elements set to one, where $V_d$ is a positive integer with $V_d<V_t$. Thus, the product $\bm a_{k,i}^T \bm e$ can remove the effect of outliers (e.g., impulsive noise samples) on the rightness of the computation of $\varDelta_k(i)$. Typically, for both $V_t$ and $V_d$, good choices are $V_t=\varrho M$ with $\varrho =1\sim3$ and $V_d=0.75V_t$ \cite{vega2008new}. Note that, for large occurence probability of impulsive noise, the value of $V_t-V_d$ should be decreased to discard the impulsive noise samples.

Secondly, if $\varDelta_k(i) > t_\text{th}$, where $t_\text{th}$ is a predefined threshold, meaning a change of $\bm w^o$ has occured, then we need to reset $\xi_k(i)$ to its initial value $\xi_k(0)$. More importantly, $\bm P_{k,i}$ is also re-initialized with $\bm P_{k,0}$. It is worth noting that since the parameters $\gamma,\;N_w,\;\varrho$, and $t_\text{th}$ are not affected by each other, their choices are simple.

\section{Simulation Results}

\begin{figure}[htb]
    \centering
    \includegraphics[scale=0.45] {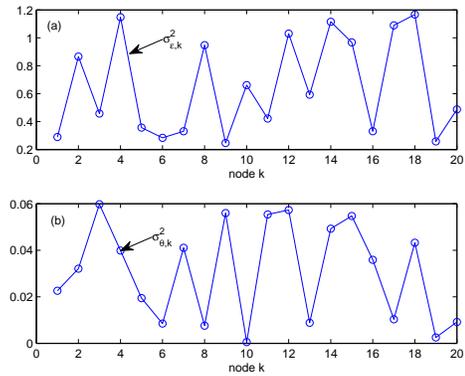}
    \hspace{2cm}\caption{Profiles of $\sigma_{\epsilon,k}^2$ and $\sigma_{\theta,k}^2$.}
    \label{Fig1}
\end{figure}

Simulation examples are presented for a diffusion network with
$N=20$ nodes. The graph describing the network is assumed to be
partially connected. Adjustments to the graph can be carried out
using approaches reported in \cite{bfpeg,dopeg,emd,vfap,kaids}. The
vector $\bm w^o$ to be estimated has a length of $M=16$ and a unit
norm; it is generated randomly from a zero-mean uniform
distribution. To evaluate the tracking capability, $\bm w^o$ changes
to $-\bm w^o$ in the middle of iterations. The input regressor $\bm
u_{k,i}$ has a shifted structure, i.e., $\bm
u_{k,i}=[u_k(i),u_k(i-1),...,u_k(i-M+1)]^T$
\cite{chouvardas2011adaptive, li2010distributed}, where $u_k(i)$ is
colored and generated by a second-order autoregressive system:
\begin{equation*}
u_k(i) = 1.6u_k(i-1)-0.81u_k(i-2)+ \epsilon_k(i),
\end{equation*}
where $\epsilon_k(i)$ is a zero-mean white Gaussian process with variance $\sigma_{\epsilon,k}^2$ shown in Fig. \ref{Fig1}(a) for all the nodes. We employ the network mean square deviation (MSD) to assess the performance of the algorithm, i.e., $\text{MSD}_\text{net}(i)=\frac{1}{N}\sum\limits_{k=1}^N E\{\|\bm w^o-\bm w_{k,i}\|_2^2\}$, where $E\{\cdot\}$ denotes the expectation. Usually, the impulsive noise can be described by either the Bernoulli-Gaussian (BG) distribution \cite{ni2016diffusion,ahn2017new,al2017robust} or the $\alpha$-Stable distribution \cite{lu2016improved, pelekanakis2014adaptive}. We consider both cases. All results are the average over 200 independent trials.

\subsection{BG Distribution}
\begin{figure}[htb]
    \centering
    \includegraphics[scale=0.53] {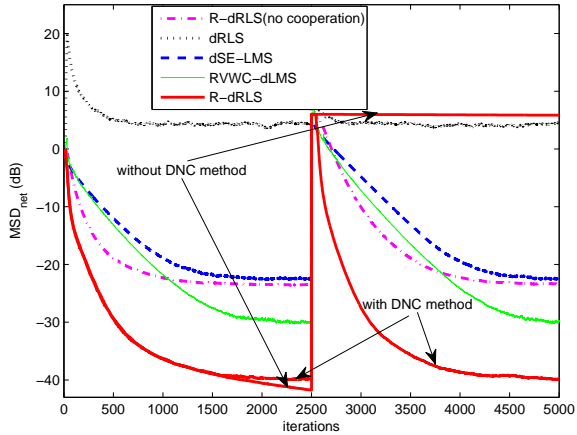}
    \hspace{2cm}\caption{MSD performance of the algorithms in impulsive noise with BG distribution. Parameter setting of the algorithms (with notations from references) is as follows: $\mu_k$=0.015 (dSE-LMS); $\beta$=0.98 and $E_c$=1 (R-dRLS); $\varrho$=3 and $t_{th}$=25 (DNC). For the RVWC-dLMS, the Metropolis rule \cite{takahashi2010diffusion} is also used for the combination coefficients in the adaptation step; its other parameters are $L$=16, $\alpha$=2.58, $\lambda$=0.98 and $\mu_k$=0.03.}
    \label{Fig2}
\end{figure}

The additive noise $v_k(i)$ includes the background noise $\theta_k(i)$ plus the impulsive noise $\eta_k(i)$, where $\theta_k(i)$ is zero-mean white Gaussian noise with variance $\sigma_{\theta,k}^2$ depicted in Fig. \ref{Fig1}(b). The impulsive noise $\eta_k(i)$ is described by the BG distribution, $\eta_k(i)=b_k(i) \cdot g_k(i)$, where $b_k(i)$ is a Bernoulli process with probability distribution $P[b_k(i)=1]=p_{r,k}$ and $\left. P[b_k(i)=0]=1-p_{r,k}\right. $, and $g_k(i)$ is a zero-mean white Gaussian process with variance $\sigma_{g,k}^2$. Here, we set $p_{r,k}$ as a random number in the range of $[0.001,0.05]$, and $\sigma_{g,k}^2=1000\sigma_{y,k}^2$, where $\sigma_{y,k}^2$ denotes the power of $y_k(i) = \bm u_{k,i}^T\bm w^o$. Fig.~\ref{Fig2} compares the performance of the dRLS, dSE-LMS, and RVWC-dLMS algorithms with that of the proposed R-dRLS algorithm. Note that, the R-dRLS (no cooperation) algorithm performs an independent estimation at each node as presented in \cite{vega2009fast}. For RLS-type algorithms, we choose $\lambda$=0.995 and $\delta$=0.01. As expected, the dRLS algorithm has a poor performance in the presence of impulsive noise. Both the dSE-LMS and RVWC-dLMS algorithms are significantly less sensitive to impulsive noise, but their convergence is slow. Apart from the robustness for combating impulsive noise, the proposed R-dRLS algorithm has also a fast convergence. Moreover, the proposed DNC method can retain the good tracking capability of the R-dRLS algorithm, only with a slight degradation in steady-state performance.

\subsection{$\alpha$-Stable Distribution}
\begin{figure}[htb]
    \centering
    \includegraphics[scale=0.53] {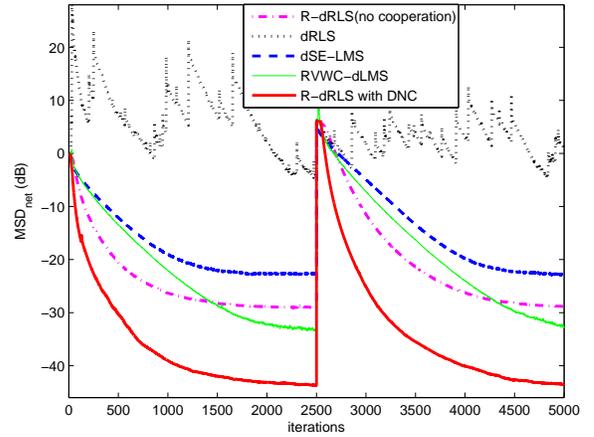}
    \hspace{2cm}\caption{MSD performance of the algorithms in $\alpha$-Stable noise. Parameter setting is the same as in Fig. \ref{Fig2}. }
    \label{Fig3}
\end{figure}

\begin{figure}[htb]
    \centering
    \includegraphics[scale=0.53] {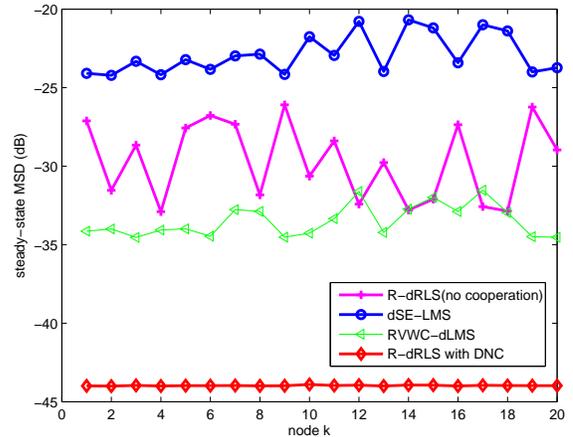}
    \hspace{2cm}\caption{Node-wise steady-state MSD of the algorithms in $\alpha$-Stable noise.}
    \label{Fig4}
\end{figure}

The impulsive noise is now modeled by the $\alpha$-stable distribution with a characteristic function $\varphi(t)=\exp(-\gamma \lvert t \lvert^\alpha)$, where the characteristic exponent $\alpha \in (0,2]$ describes the impulsiveness of the noise (smaller $\alpha$ leads to more impulsive noise samples) and $\gamma>0$ represents the dispersion level of the noise. In particular, when $\alpha=2$, it reduces to the Gaussian noise. It is rare to find $\alpha$-stable noise with $\alpha<1$ in practice \cite{pelekanakis2014adaptive,shao1993signal}. In this example, thus we set $\alpha=1.15$ and $\gamma=1/15$. The learning performance of the algorithms is shown in Fig.~\ref{Fig3}. Fig. \ref{Fig4} shows the node-wise steady-state MSD of the robust algorithms (i.e., excluding the dRLS) against impulsive noise, by averaging over 500 instantaneous MSD values in the steady-state. As can be seen from Figs. \ref{Fig3} and \ref{Fig4}, the proposed R-dRLS algorithm with DNC outperforms the known robust algorithms.

\section{Conclusion}

In this paper, the R-dRLS algorithm has been proposed, based on the
minimization of an individual RLS cost function with a
time-dependent constraint on the squared norm of the intermediate
estimate update. The constraint is dynamically adjusted based on the
diffusion strategy with the help of side information. The novel
algorithm not only is robust against impulsive noise, but also has
fast convergence. Furthermore, to track the change of parameters of
interest, a detection method (DNC method) is proposed for
re-initializing the constraint. Simulation results have verified
that the proposed algorithm performs better than known algorithms in
impulsive noise scenarios.
\newpage

\bibliographystyle{IEEEbib}
\bibliography{mybib}

\end{document}